\documentstyle[latexsym,amssymb,aps,floats,eqsecnum,preprint,amsfonts]{revtex}
\def\laq{\raise 0.4ex\hbox{$<$}\kern -0.8em\lower 0.62 ex\hbox{$\sim$}}
\def\gaq{\raise 0.4ex\hbox{$>$}\kern -0.7em\lower 0.62 ex\hbox{$\sim$}}
\input epsf
\begin{document}
\draft
\bibliographystyle{unsrt}

\title{{Six-dimensional Abelian vortices}\\{ with quadratic curvature 
self-interactions}}

\author{M. Giovannini and  H. B.  Meyer}

\address{{\it Institute of Theoretical Physics, 
University of Lausanne}}
\address{{\it BSP-1015 Dorigny, Lausanne, Switzerland}}

\maketitle

\begin{abstract} 
Six-dimensional Nielsen-Olesen vortices are analyzed in the 
context of a quadratic gravity
theory  containing Euler-Gauss-Bonnet
self-interactions. The relations among the string tensions can be tuned 
in such a way that the obtained solutions lead to warped compactification 
on the vortex. New regular solutions are possible in comparison 
with the case where the gravity action only consists 
of the Einstein-Hilbert term. The parameter space of the model is discussed.
\end{abstract}
\vskip0.5pc
\centerline{Preprint Number: UNIL-IPT-01-12, August 2001 }
\vskip0.5pc
\noindent

\newpage
\renewcommand{\theequation}{1.\arabic{equation}}
\setcounter{equation}{0}
\section{Introduction} 

Higher dimensional gravity theories have been discussed in 
connection with possible alternatives to Kaluza-Klein compactification 
\cite{m1,m2,ak,vis,rs,rs2}. In this context the gravity part of the  
action usually consists of the Einstein-Hilbert 
term possibly supplemented by a bulk cosmological constant. 

If extra-dimensions are not compact, it is 
interesting to relax the assumption that gravity 
is described purely in terms of the Einstein-Hilbert action. Indeed, 
various investigations took into account the 
possible contribution of higher derivatives terms in the bulk action
 and mainly 
in the case of one bulk coordinate \cite{q1,q2,q3,q4,q5}. The motivations 
of these studies range from the possible connection to string 
motivated scenarios \cite{q1} to the possibility of obtaining 
warped compactifications in quadratic gravity theories \cite{lov,mad}
defined in more than four dimensions. 
Quadratic corrections may also play a r\^ole in the context of AdS/CFT 
correspondence \cite{cft}.

Recently, it has been  shown that
gravity can be localized on a Nielsen-Olesen vortex in the context 
of the Abelian-Higgs model \cite{us} in a six dimensional space-time.
This explicit model leads to warped compactification and it 
represents an explicit field theoretical realization of the suggestion
that a string in six dimensions can lead to the localization 
of gravity \cite{gs} provided certain relations among the 
string tensions are satisfied. In the investigations dealing with 
higher dimensional topological defects (like strings 
\cite{def,def1,def2,def3,def4}, 
monopoles \cite{Dvali:2000ty,grs,Randjbar-Daemi:1983qa}, 
or instantons \cite{Randjbar-Daemi:2000cr,Randjbar-Daemi:2000ft,gre})  
the space time metric is defined in more than five dimensions, but, still,
the gravity part 
of the action is assumed to be the Einstein-Hilbert 
term possibly supplemented by a bulk cosmological constant 
\footnote{Notice that the 
present investigation deals with the case when the internal dimensions 
are not compact. The interesting case of compact (but large) extra-dimensions 
\cite{dim,dim1,dim2} will then be outside the scope of this analysis. }.
 
We would like to understand if it is possible to obtain warped 
solutions in the context of the gravitating Abelian-Higgs model in six 
dimensions when quadratic curvature corrections are present. 
The brane action will be the one provided by the Abelian-Higgs model 
\cite{no} generalized to the case of six (gravitating) dimensions.
The quadratic corrections to the Einstein-Hilbert action will be 
taken in the Euler-Gauss-Bonnet (EGB) form. 
This choice has the interesting 
property of preserving the relations among the string tensions derived 
in the case when the gravity action is in the Einstein-Hilbert form.

Previous  studies suggest that if the Einstein Hilbert action is 
supplemented by quadratic curvature corrections, 
geometries leading to warped compactification in more than 
five dimensions can be obtained.
More specifically, in \cite{mg1} seven-dimensional 
warped solutions have been discussed in the case when 
hedgehogs configurations are present together with 
quadratic self-interactions parametrized 
in the Euler-Gauss-Bonnet form. In \cite{mg2} the simultaneous presence of 
dilaton field and quadratic corrections has been investigated in a 
string theoretical perspective and mainly in six-dimensions.

The plan of this paper is the following. In Section II the main equations 
of the system are derived. Section III is devoted to the analysis 
of the asymptotics of the solutions in the presence of EGB corrections. 
Section IV contains the study of the relations among the string tensions,
whereas Section V deals with the 
analysis of the various classes of solutions which can be 
found in the presence of quadratic corrections. 
In Section VI the analysis of the parameter space of the model is presented.
Section VII contains  our concluding remarks. 

\renewcommand{\theequation}{2.\arabic{equation}}
\setcounter{equation}{0}
\section{Brane sources with quadratic gravity in the bulk} 

In more than four dimensions the Einstein-Hilbert term
is not the only geometrical action leading to equations of motion 
involving  second order derivatives of the metric \cite{lov}.
The usual Einstein-Hilbert invariant can 
indeed be supplemented with higher order curvature corrections 
without generating, in the equations of motion, terms containing 
more than two derivatives of the metric with respect to the 
space-time coordinates \cite{mad}. If this is the case, the quadratic part 
of the action can be written in terms of the Euler-Gauss-Bonnet 
combination:
\begin{equation}
{\cal R}^2_{\rm EGB} = R^{ A B C D} R_{A B C D} - 4 R^{A B} R_{A B} + R^2.
\label{egb}
\end{equation}
In four dimensions the EGB is a topological term and it coincides with 
the Euler invariant: its contribution to the equations of 
motion can be rearranged in a perfect four-divergence which does not
contribute to the classical equations of motion. In more than 
four dimensions the EGB combination leads to a ghost-free 
theory and it appears in different higher 
dimensional contexts. In string 
theory the (tree-level) low energy effective action is 
supplemented not only by an expansion in the powers of 
the dilaton coupling but also by an expansion in powers of the 
string tension $\alpha'$. The EGB indeed appears in the first 
$\alpha'$ correction \cite{des,ts,cal,s}.
In supergravity the EGB is required in order to supersymmetrize the 
Lorentz-Chern-Simons term. 

The brane source will be described 
in terms of the Abelian-Higgs model \cite{no} appropriately 
generalized to the six-dimensional case \cite{us}:
\begin{equation}
S_{\rm brane}=\int
d^6x\sqrt{-G}\biggl[
\frac{1}{2}({\cal D}_{A}\phi)^*{\cal D}^{A}\phi-\frac{1}{4}
F_{AB}F^{AB}
-\frac{\lambda}{4}\left(\phi^*\phi-v^2\right)^2\biggr]~,
\label{bac}
\end{equation}
where ${\cal D}_{A}=\nabla_{A}-ieA_{A}$ is the gauge covariant derivative, 
while $\nabla_{A}$ is the generally covariant derivative 
\footnote{The conventions 
of the present paper are the following : the signature 
of the metric is mostly minus, Latin (uppercase) indices run over 
the $(4+2)$-dimensional space. Greek indices run over the four-dimensional 
space-time.}.

The bulk action will 
then be taken  in the form
\begin{equation}
S_{\rm bulk} = -\int d^{6} x \sqrt{- G} \biggl[ \frac{R}{2\chi } + \Lambda 
- \alpha' {\cal R}^2_{\rm EGB} \biggr].
\label{acbul}
\end{equation}
This  gravitational effective action can be related to the 
low energy string effective action (corrected to first order in the 
string tension) provided the dilaton and antisymmetric tensor field 
are frozen to a constant value \cite{ts}. Since 
$ \chi = 8 \pi G_{6} = 8\pi/ M_{6}^4$, we have that, dimensionally, 
$\alpha' = [M_{6}]^2$.

With these conventions, the equations of motion of the system can be obtained
and  they are:
\begin{eqnarray}
&&G^{A B}\nabla_{A}\nabla_{B}\phi
-e^2A_{A}A^{A}\phi-ieA_{A}\partial^{A}\phi-ie\nabla_{A}(A^{A}\phi)
+\lambda (\phi^*\phi-v^2)\phi=0,
\label{ph}\\
&&\nabla_{A} F^{AB}=-e^2A^{B}\phi^*\phi+\frac{ie}{2}
\left(\phi\partial^{B}\phi^*-\phi^*\partial^{B}\phi\right),
\label{A}\\
&& R_{AB}-\frac{1}{2}G_{AB}R = \chi\left(T_{AB}+\Lambda G_{AB}\right)  -
2 \alpha' \chi {\cal Q}_{AB} ,
\label{R}
\end{eqnarray}
where 
\begin{eqnarray}
T_{AB}&=& \biggl[
\frac{1}{2}({\cal D}_{M}\phi)^*{\cal D}^{M}\phi-\frac{1}{4}
F_{MN}F^{MN}
-\frac{\lambda}{4}\left(\phi^*\phi-v^2\right)^2\biggr] 
G_{AB}\,
\nonumber\\
&+&\, \frac{1}{2}\left[({\cal D}_A\phi)^*{\cal D}_{B}\phi 
+({\cal D}_B\phi)^*{\cal D}_A\phi\right]
-\,F_{AC}{F_B}^{C},
\end{eqnarray}
and where 
\begin{equation}
{\cal Q}_{A B} = \frac{1}{2} G_{A B} {\cal R}_{\rm EGB}^2 - 2 R R_{AB} + 
4 R_{A C} R^{~~C}_{B} + 4 R_{CD} R_{A~B}^{~~C~~D} - 2 R_{A C D E}R_{B}^{~~CDE}
\end{equation}
is the Lanczos tensor appearing as a result of the  EGB contribution in the 
action.

Defining as $n$ the winding number, the Nielsen-Olesen ansatz \cite{no}
can be generalized to six-dimensional space-time 
\begin{eqnarray}
&&\phi(\rho,\theta) = v f(\rho) e^{ i\, n\, \theta}, 
\nonumber\\
&&A_{\theta}(\rho,\theta)  =\frac{1}{e}[\,n\,-\,P(\rho)] ~,
\label{NO}
\end{eqnarray}
where $\rho$ and $\theta$ are, respectively, the bulk radius and the 
bulk angle covering the transverse space of 
the metric 
\begin{equation}
ds^2=G_{AB} dx^{A} dx^{B} =
M^2(\rho)g_{\mu\nu}dx^\mu dx^\nu-d\rho^2-L(\rho)^2d\theta^2.
\label{metric}
\end{equation}
In Eq. (\ref{metric}), we will choose, for simplicity
 $g_{\mu\nu}\equiv \eta_{\mu\nu}$ where $\eta_{\mu\nu}$ is the Minkowski 
metric.

The equations of motion for the gauge-Higgs system can then be written.
Using Eqs. (\ref{NO})--(\ref{metric}) into Eqs. (\ref{ph})--(\ref{R})
the following system of equations is obtained:
\begin{eqnarray} 
&& f'' + ( 4 m + \ell ) f'+(1 - f^2) f -
\frac{P^2}{{\cal L}^2}f=0,
\label{f1}\\ 
&& P'' + ( 4 m - \ell) P'  -\alpha f^2 P=0,
\label{p1}\\ 
&& \ell' + 3 m' + \ell^2 + 6 m^2 + 3 l m  = - \mu - \nu \tau_0 
+\epsilon {\cal G}_0,
\label{m1}\\
&& 4 m' + 10 m^2 = - \mu - \nu \tau_{\theta}
 + \epsilon {\cal G}_{\theta},
\label{m2}\\
&& 4 m \ell + 6m^2 = -\mu - \nu \tau_{\rho}
+\epsilon {\cal G}_{\rho}.
\label{l1}
\end{eqnarray}
Eqs. (\ref{f1}) and (\ref{p1}) correspond, respectively, to the 
evolution equations for the Higgs field and for the gauge field. 
The remaining equations come from the various components 
of Eq. (\ref{R}). In Eqs. (\ref{m1})--(\ref{l1}) the 
 non-vanishing components of the 
energy-momentum tensor are
\begin{eqnarray} 
\tau_0(x)  &\equiv& T_{0}^{0}= T_{i}^{i}
= \frac{ {f'}^2}{2} + \frac{1}{4} ( f^2 -1 )^2 
+ \frac{{P'}^2}{2 \alpha {\cal L}^2 } + \frac{f^2 P^2 }{2
 {\cal L}^2},
\label{t0}\\ 
\tau_\rho(x)&\equiv& T_{\rho}^{\rho} =-\frac{{f'}^2}{2} 
+ \frac{1}{4} ( f^2 -1 )^2 - \frac{{P'}^2}{2 \alpha {\cal L}^2 } 
+ \frac{f^2 P^2 }{2 {\cal L}^2},
\label{trho}\\ 
\tau_\theta(x)&\equiv& T_{\theta}^{\theta}=
 \frac{{f'}^2}{2} + \frac{1}{4}  ( f^2 -1 )^2 
- \frac{{P'}^2}{2 \alpha {\cal L}^2 } - \frac{f^2 P^2  }{2 {\cal L}^2}.
\label{tth} 
\end{eqnarray}
Finally, 
the functions ${\cal G}$ appearing in Eqs. (\ref{m1})--(\ref{l1}) 
are the non-vanishing components of the Lanczos 
tensor ${\cal Q}_{A B}$:
\begin{eqnarray}
{\cal G}_0&=& 36\ell m^3 + 12 m^4 + 12 m^2 \ell' + 12 m^2 \ell^2 +  
+24\ell mm'+12m^2m',
\label{G0}\\
{\cal G}_\rho&=&12m^4+48m^3\ell, 
\label{Gr}\\
{\cal G}_\theta&=&48m^2m'+60m^4.
\label{Gth}
\end{eqnarray}
Eqs. (\ref{f1})--(\ref{l1}) are written in terms 
of the rescaled parameters:
\begin{equation} 
{\cal L}(x) = \frac{m_{H}}{\sqrt{2}} L(\rho),~~~
\alpha= 2 \frac{m^2_{V}}{m^2_{H}},~~~
\mu =\frac{2 \chi \Lambda}{ m^2_{H}},~~~ 
\nu= \chi \frac{m^2_{H}}{2\lambda},
\label{constant}
\end{equation}
where $ m_{H} = \sqrt{2 \lambda} v$ is the Higgs mass and 
$m_{V} = e ~v $ is the vector boson mass. In Eq. (\ref{f1})--(\ref{l1}) 
the prime denotes the derivative with respect to $x = m_{H}\rho/\sqrt{2}$ (the 
rescaled bulk radius) and $m(x) = M'/M $ , $\ell(x) = {\cal L}'/{\cal L}$.
In Eqs. (\ref{f1})--(\ref{l1}) 
the $\epsilon =  \alpha' \chi m_{H}^2 
$ has been defined. This parameter 
controls the higher derivative corrections.

\renewcommand{\theequation}{3.\arabic{equation}}
\setcounter{equation}{0}
\section{Asymptotics of the solutions} 
In order to discuss regular solutions describing a thick brane in six 
dimensions and in the presence of EGB self-interactions 
it is important to understand the asymptotic behaviour of the geometry 
and of the matter fields of the model. Quadratic curvature corrections 
affect the asymptotics of the geometry and the asymptotics of the scalar
and gauge fields. 
\subsection{Bulk solutions}
Far from the core of the defect,  
the equations determining the bulk solutions are given by
\begin{eqnarray}
&& m' = - \frac{5}{2} m^2 - \frac{ \mu + 60 \epsilon m^4}{4( 1 - 12 \epsilon 
m^2)},
\label{mbul}\\
&& \ell = - \frac{m}{4} - \frac{\mu + 5 m^2}{4 m ( 1 - 12 \epsilon m^2)}.
\label{lbul}
\end{eqnarray}
From Eq. (\ref{mbul}), $m(x)$ can be determined and Eq. (\ref{lbul}) will 
then give $\ell(x)$. 
If $m'(x) =0$ and $\ell'(x)=0$ a consistent solution can be found 
provided  $\ell(x) = m(x)$. In this case $m$ should satisfy the 
following quartic equation
\begin{equation}
 60 \epsilon m^4 - 10 m^2 - \mu =0.
\label{detm}
\end{equation}
whose solution  can be written as
\begin{equation}
m^2 = \frac{1}{12 \epsilon} 
\biggl( 1 \pm \sqrt{ 1 + \frac{12}{5} \epsilon\mu}\biggr).
\end{equation}
Defining now
\begin{eqnarray}
&&\gamma_{1} = \frac{1}{12\epsilon} \biggl( 1 +
 \sqrt{ 1 + \frac{12}{5} \epsilon\mu}\biggr),
\nonumber\\
&& \gamma_{2} = \frac{1}{12 \epsilon} 
\biggl( 1 - \sqrt{ 1 + \frac{12}{5} \epsilon\mu}\biggr),
\end{eqnarray}
warped solutions will correspond either to 
 $m_{1} \sim - \sqrt{\gamma_1} $ or to 
$m_2 \sim - \sqrt{\gamma_2}$. As a consequence, the behaviour of 
the metric functions will be either 
\begin{equation}
M(x) \sim M_0 e^{ - \sqrt{\gamma_1} x}, \,\,\, {\cal L}(x) \sim {\cal L}_0 
e^{- \sqrt{\gamma_1} x},
\label{as1}
\end{equation}
or   
\begin{equation}
M(x) \sim M_0 e^{ - \sqrt{\gamma_2} x}, \,\,\, {\cal L}(x) \sim {\cal L}_0 
e^{- \sqrt{\gamma_2} x}.
\label{as2}
\end{equation}
Since $\gamma_1$ and $\gamma_2$ must both be positive 
for warped solutions, not all the signs of $\epsilon$ and $\mu$ are compatible
with the behaviour of Eqs. (\ref{as1})--(\ref{as2}).
More precisely, 
if $\epsilon > 0$ and $\mu>0$ only the solution $\gamma_1$ is possible. 
If $\epsilon <0$  and $\mu <0$ only the solution $\gamma_2$ is possible. 
If $\epsilon>0$ and $\mu <0$  both, $\gamma_1$ and $\gamma_2$, 
lead to warped geometries. 
Finally, if $\mu>0 $ and $\epsilon <0$ no solution is possible. 
The situation can be summarized in the following table.
\begin{center}
\begin{tabular}{|c||c|c|}
\hline
 & $\epsilon>0$ &  $\epsilon<0$ \\
\hline \hline
 $\mu>0$ & $\gamma_1$ & $\emptyset$ \\
\hline
 $\mu<0$ & $\gamma_1,~\gamma_2$ & $\gamma_2$ \\
\hline
\end{tabular}
\end{center}
Let us now investigate what happens if $m(x) \neq \ell(x)$.
Subtracting Eq. (\ref{l1}) from Eq. (\ref{m2}) we obtain 
\begin{equation}
( 1 - 12 \epsilon m^2)[m' - m( \ell - m)] =0,
\label{crit}
\end{equation}
If $m'(x)=0$ and $\ell'(x) =0$,
Eq. (\ref{crit}) can be satisfied in two different ways. If 
$m(x) = -1/\sqrt{12\epsilon}$, then Eq. (\ref{crit}) can be  solved, in 
principle,
with $\ell(x) \neq m(x)$. However, $m(x) = -1/\sqrt{12\epsilon}$ is 
not consistent \footnote{ Notice that a similar kind of 
argument will be used in Section IV in order to show that the 
solutions with positive cosmological constant lead to singular space-times.}  
with Eq. (\ref{mbul}).
Hence, the only relevant solutions, in the present context, are the ones 
for which $m(x) \neq -1/\sqrt{12\epsilon}$ and $\ell(x) = m(x)$.

\subsection{Asymptotics of vortex solutions with quadratic gravity}

The system defined by  equations (\ref{f1})--(\ref{l1})
describes the interactions of the bulk degrees of freedom with 
the brane degrees of freedom. The problem will be completely 
specified only when specific boundary conditions for the fields 
will be required. In order to describe a string-like defect in six 
dimensions it has to be demanded that the scalar field 
reaches, for large bulk radius, its vacuum expectation 
value, namely $|\phi(\rho)| \rightarrow v$ 
for $\rho \rightarrow \infty$. In the same limit, the magnetic field
should go to zero. Moreover, close to the core of the string 
both fields should be regular. These requirements 
can be translated in terms of our 
rescaled variables using the generalized Nielsen-Olesen vortex of 
Eq. (\ref{NO}):
\begin{eqnarray} 
f(0)=0,&\qquad& \lim_{x\rightarrow \infty} f(x)=1,
\nonumber\\ 
P(0)=n,&\qquad& \lim_{x\rightarrow \infty} P(x)=0.
\label{boundary}
\end{eqnarray}
In the following part of the analysis we will be concerned with the case 
of lowest winding, i.e. $n =1$. Solutions for higher windings 
can be however discussed using the same results of the present analysis 
and they may be relevant for the problem of fermion localization \cite{tro}.

In order to generalize vortex-type solutions to the 
case of curvature self-interactions we have to demand that 
the boundary conditions expressed by Eqs. (\ref{boundary}) are satisfied 
by Eqs. (\ref{f1})--(\ref{l1}) which are different from the case 
where the gravity action only contains the Einstein-Hilbert term and 
the bulk cosmological constant.
Close to the core of the vortex, the consistency 
of Eq. (\ref{f1})--(\ref{l1}) with the vortex ansatz (\ref{NO}) 
satisfying the boundary conditions (\ref{boundary}) requires that 
\begin{eqnarray}
f(x)&=&Ax+\frac{A}{8}\left[\frac{2\mu}{3}+\frac{\nu}{6}
+\frac{4\nu B^2}{3\alpha}
+\frac{2\nu A^2}{3}-1+2B-\epsilon\left(\mu+\frac{\nu}{4}-\frac{2\nu 
B^2}{\alpha}\right)^2 \right]x^3,\\
P(x)&=&1+Bx^2,\\
M(x)&=&1-\left(\frac{\mu}{8}+\frac{\nu}{32}-\frac{\nu 
B^2}{4\alpha}\right)x^2,\\
L(x)&=&x+\left[\frac{\mu}{12}+\frac{\nu}{48}-\frac{5\nu B^2}{6\alpha}-
\frac{\nu 
A^2}{6}+\epsilon\left(\frac{\mu}{2}+\frac{\nu}{8}-\frac{\nu B^2}{\alpha} 
\right)
^2  \right]x^3,
\label{small}
\end{eqnarray}
The small argument limit of the 
solution explicitly depends upon $\epsilon$. 

\renewcommand{\theequation}{4.\arabic{equation}}
\setcounter{equation}{0}
\section{Generalized Relations among the string tensions} 

In Eq. (\ref{small}) on top of the physical parameters of the 
model (i.e. $\mu$, $\nu$, $\alpha$ and $\epsilon$ ) the constants $A$ and $B$ 
appear. These constants 
 are arbitrary and should be fixed in order to match the correct
boundary conditions at infinity. In the case where the curvature corrections
are absent one of the two constants, i.e. $B$, can be fixed by looking at the
relations obeyed by the string tensions. The other constant, i.e. $A$ 
cannot be fixed from the relations among the string tensions and it 
represents a free parameter of the numerical integration.
According to the results of \cite{us} the possibility of 
fixing $B$ through the relations among the string tensions 
is {\em necessary}. 
This means that in order to hope that 
warped solutions leading to gravity localization are present  
in the presence of EGB corrections, we should be able 
to generalize the relations among the various string tensions 
which are defined as 
\begin{equation}
\mu_i = \int_0^\infty dx M^4(x){\cal L}(x)\tau_i(x).
\label{tension}
\end{equation}
The relations connecting different $\mu_i$ are integral relations 
and connect, therefore, the behaviour of the solution 
close to the core with the behaviour of the solutions at infinity.

Formally we can also define the integrated components of 
the Lanczos tensor, namely
\begin{equation}
g_{i} =  \int_0^\infty dx M^4(x){\cal L}(x){\cal G}_i(x).
\end{equation}
Combining Eqs. (\ref{f1})--(\ref{l1}) the following 
relations can be obtained:
\begin{eqnarray}
&&2m' + 4 m^2 + m\ell =-\frac{\mu}{2}-\frac{\nu}{4}\left[  
\tau_\rho
+ \tau_\theta \right] + \frac{\epsilon}{4} 
[ {\cal G}_{\rho} + {\cal G}_{\theta} ],
\label{c1}\\
&& \ell' + \ell^2 + 4\ell m 
=-\frac{\mu}{2}-\nu\left[  \tau_0+\frac{1}{4} \tau_\rho
-\frac{3}{4} \tau_\theta \right] +\epsilon\biggl[ {\cal G}_{0} + \frac{1}{4} 
{\cal G}_{\rho} - \frac{3}{4} {\cal G}_{\theta} \biggr].
\label{c2}
\end{eqnarray}
Integrating Eqs. (\ref{c1})--(\ref{c2}) from zero to $x_c \to \infty$, 
we get
\begin{eqnarray}
&& M^3(x_c)M'(x_c){\cal L}(x_c)=-\frac{\mu}{2}\int_0^{x_c} 
M^4 {\cal L}dx
-\frac{\nu}{4}\left(\mu_\rho+\mu_{\theta}\right) 
+ \frac{\epsilon}{4} (g_{\rho} + g_{\theta}) ,
\label{rela}\\ 
&& M(x_c)^4{\cal L}'(x_c)=1-\frac{\mu}{2}\int_0^{x_c} M^4 {\cal L}dx-\nu
\left(\mu_0-\frac{3}{4}\mu_{\theta}
+\frac{\mu_\rho}{4}\right) +\epsilon \biggl( g_{0} - \frac{3}{4} 
g_{\theta} + \frac{1}{4} g_{\rho}
\biggr).
\label{relb} 
\end{eqnarray} 
In the limit $x_c \to \infty$, Eq. (\ref{rela}) is the 
six-dimensional analog  of the relation determining the Tolman mass
in the presence of EGB corrections.
Eq. (\ref{relb})  is the generalization of the relation
giving the angular deficit to the case of a six-dimensional 
geometry sustained by quadratic corrections.
Hence, the relation 
\begin{equation}
\mu_{0} - \mu_{\theta} = \frac{1}{\nu} + \frac{\epsilon}{\nu} 
( g_0 - g_{\theta}) 
\label{gennu}
\end{equation}
must hold.
If we now notice that 
\begin{equation}
g_{0} - g_{\theta} = \int_0^{\infty} 
M^4 {\cal L} ({\cal G}_0 - {\cal G}_{\theta} ) dx
\equiv 12 \int_{0}^{\infty} \biggl( M^4 {\cal L} m^2 ( \ell - m)\biggl)' dx,
\end{equation}
we have that 
\begin{equation}
g_{0} - g_{\theta} = 12 \biggl(\left. M  {M'}^2( {\cal L}' M - M' {\cal L})
\right|_{\infty} -\left.  M  {M'}^2( {\cal L}' M - M' {\cal L})
\right|_{0}\biggr).
\label{gt}
\end{equation}
Notice now that Eqs. (\ref{f1}) and (\ref{p1}) can be used 
in order to express $\mu_{0} - \mu_{\theta}$ in terms of the physical fields 
appearing in the energy-momentum tensor.
More specifically, using Eq. (\ref{p1}) in the form 
\begin{equation}
\left(\frac{M^4P'}{{\cal L}}\right)'=\alpha\frac{f^2PM^4}{\cal L},
\end{equation}
we also have that
\begin{equation}
\mu_0-\mu_\theta=\frac{1}{\alpha}\biggl(
\left.\frac{P\,P'\,M^4}{\cal L}\right|_{\infty}
-\left.\frac{P\,P'\,M^4}{\cal L}\right|_0\biggr)~.
\label{in2}
\end{equation}
For the solutions we are interested in, 
the boundary term at infinity vanishes. Therefore  we have 
\begin{equation}
\mu_0-\mu_\theta=
-\frac{1}{\alpha}\left.\frac{P'}{\cal L}\right|_0.
\label{in3}
\end{equation}
Inserting Eqs. (\ref{gt}) and (\ref{in2}) into Eq. (\ref{gennu}) and using 
Eq. (\ref{in3}) we get 
\begin{equation}
-\frac{\nu}{\alpha}\left.\frac{P'}{\cal L}\right|_0=1\,.
\label{mus2}
\end{equation}
According to Eq. (\ref{small}), for $x \rightarrow 0$, 
$ P\sim 1 + B x^2$. Using 
Eq. (\ref{mus2}) the expression for $B$ can be exactly computed
\begin{equation}
B= - \frac{\alpha}{2\nu}\,.
\label{B}
\end{equation}
This result coincides with the result obtained without 
quadratic corrections even if quadratic corrections 
do contribute to the equations of motion 
in a non-trivial way. The reason of this 
result is that Gauss-Bonnet self-interaction have the 
property of preserving some of the relations among the 
string tensions.

This aspect can be better appreciated using the following observation.
Consider  the difference between the $(0,0)$ and $(\theta,\theta)$
components of the corrected Einstein equations, i.e. (\ref{m1}) and (\ref{m2}):
\begin{equation}
(\ell' - m') + ( \ell + 4 m) (\ell - m) = - \nu( \tau_0 - \tau_{\theta} ) 
+ \epsilon\, ({\cal G}_{0} - {\cal G}_{\theta}). 
\end{equation}
Multiplying both sides of this equation by $M^4 {\cal L}$ and integrating 
from $0$ to $x$ we get 
\begin{equation}
( m - \ell) ( 1 - 12 \epsilon m^2) = \frac{\nu}{\alpha} 
\frac{ P P'}{{\cal L}^2} - \biggl( 1 + \left.
 \frac{\nu}{\alpha} \frac{P'}{{\cal L}}
\right|_0\biggr) \frac{1}{M^4 {\cal L}}.
\label{tuneq}
\end{equation}
If the tuning among the string tensions is enforced, the boundary term 
in the core disappears and the resulting equation will be 
\begin{equation}
\ell = m - \frac{\nu}{\alpha} \frac{P~ P'}{{\cal L}^2 ( 1 -12 \epsilon m^2)}.
\end{equation}
This equation shows, again, that if $P' \rightarrow 0$ (far from the core), 
we must have, asymptotically, that $m(x) \simeq \ell(x)$.

\renewcommand{\theequation}{5.\arabic{equation}}
\setcounter{equation}{0}
\section{Vortex solutions with Gauss-Bonnet self-interactions} 

In order to study first analytically and then numerically the 
full system of Eqs. (\ref{f1})--(\ref{m1}) with the asymptotics 
of a generalized Nielsen-Olesen vortex it is useful to 
recast the equations for $\ell$ and $m$ is the form of a dynamical system. 
This reduction is still possible in the case of 
curvature corrections since the EGB term does not introduce, in the 
equations of motion, terms like $m''$ or $m'''$. If the bulk 
cosmological constant is negative warped solutions are possible 
both for $\epsilon >0$ and $\epsilon<0$. For theoretical 
reasons the case $\epsilon >0$ is more physical \cite{des,des,ts}.

More specifically, if  $\mu <0$ and in the absence of quadratic 
corrections there are
geometries leading to singular curvature invariants. If quadratic 
corrections are included in the EGB form these solutions
become regular and exponentially decreasing. 
On the other hand, if the bulk cosmological 
constant is positive (i.e. $\mu >0$ ) exponentially decreasing solutions 
are never possible even in the presence of the EGB combination. In the 
following part of this Section these two cases will be specifically analyzed.

\subsection{Solutions with $\Lambda < 0$}

If the cosmological constant is negative warped solutions can be found. Here
we are going to compare solutions with and without quadratic corrections 
obeying fine tuned relations among the string tensions. In the previous 
Section we argued that a necessary condition in order to have 
warped geometries leading to gravity localization is represented 
by Eq. (\ref{B}).

\begin{figure}
\centerline{\epsfxsize = 11 cm  \epsffile{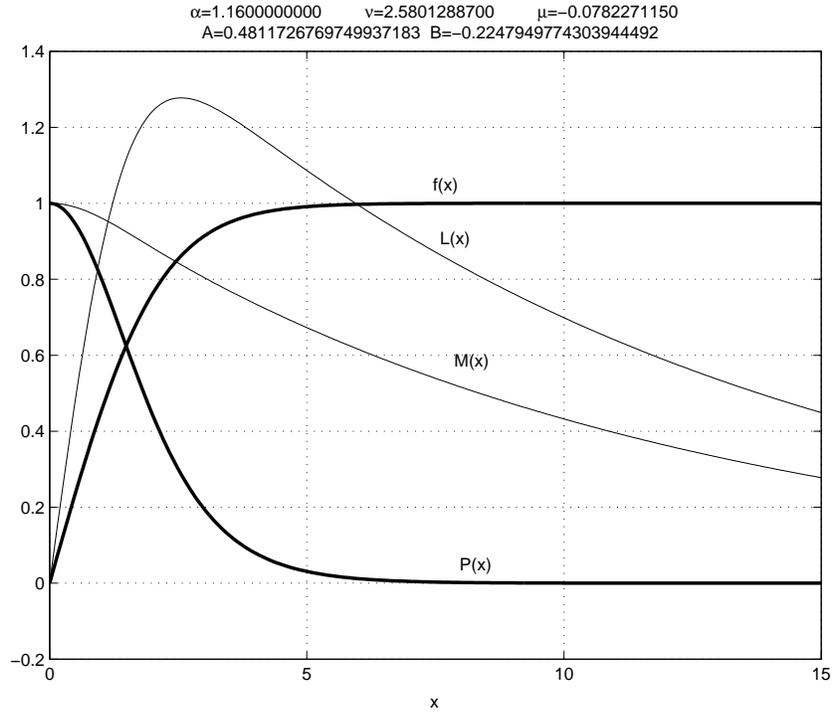}} 
\caption[a]{A vortex-like solution of the Abelian-Higgs model leading 
to gravity localization.}
\label{F1}
\end{figure}

\begin{figure}
\centerline{\epsfxsize = 11 cm  \epsffile{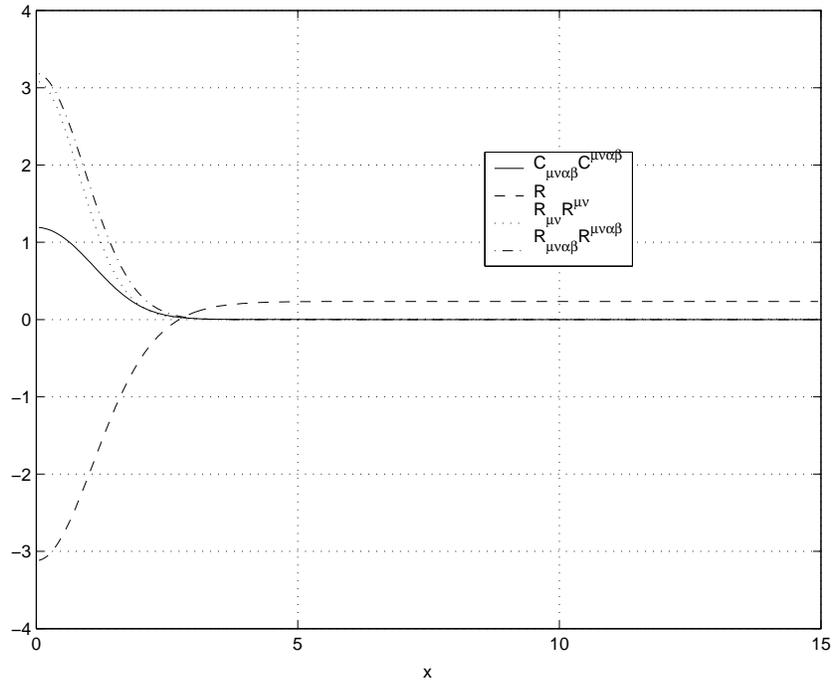}} 
\caption[a]{The curvature invariants in the case of the solution 
given in Fig. \ref{F1}.}
\label{F2}
\end{figure}

\begin{figure}
\centerline{\epsfxsize = 11 cm  \epsffile{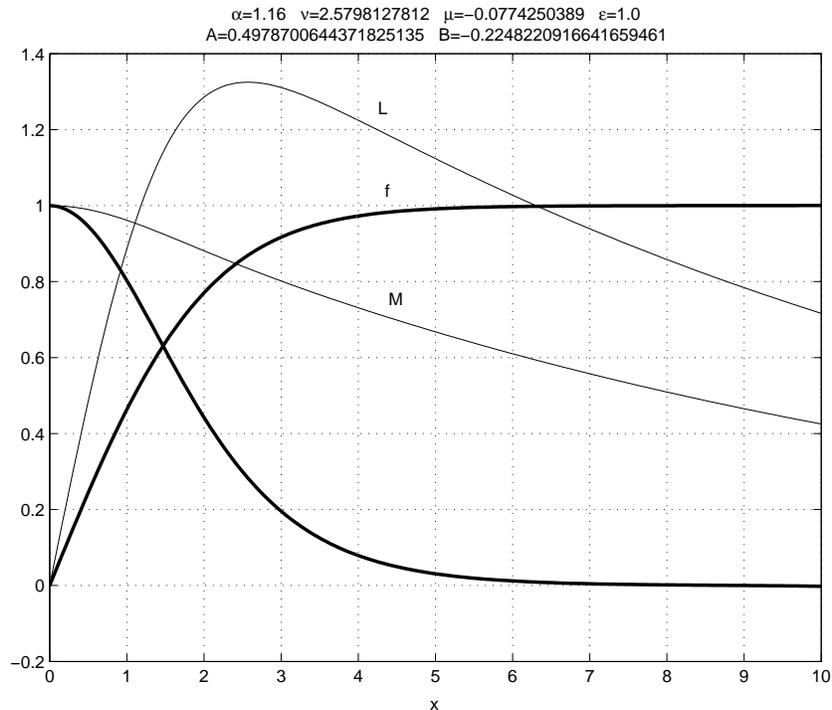}} 
\caption[a]{A regular solution for $\epsilon >0$ is illustrated. 
The parameters of the solution are the ones 
of Fig. \ref{F1} but adapted to the new value of $\epsilon$, i.e. 
 $\epsilon =1$.}
\label{F3}
\end{figure}

\begin{figure}
\centerline{\epsfxsize = 11 cm  \epsffile{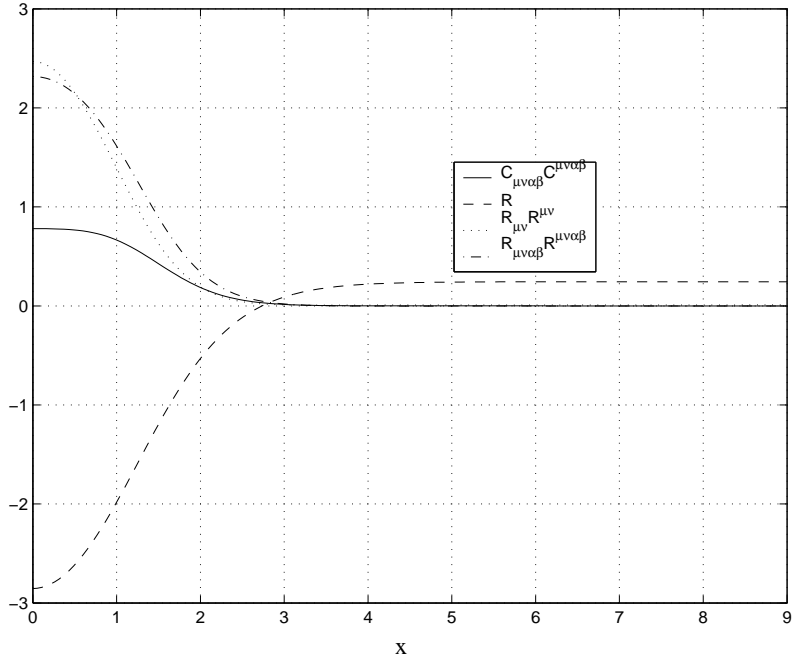}} 
\caption[a]{The curvature invariants in the case of the solution 
given in Fig. \ref{F3}.}
\label{F4}
\end{figure}

Now the exercise will be the following. Take a regular 
solution in the absence of curvature corrections and see what happens
by switching on the contribution of the Lanzos tensor. 
Consider, first, a solution which is regular in the absence 
of curvature corrections. This solutions corresponds to the case 
$\epsilon=0$. In this case 
the solution and the behaviour of the curvature invariants is illustrated, 
respectively, in Figs. \ref{F1} and  \ref{F2}.
Consider now the parameter set leading to 
the regular solution reported in Fig. \ref{F1} where $\epsilon =0$.
If we now change continuously $\epsilon$ from   
$\epsilon =0$ to $\epsilon > 0$ new solutions (still regular) can be found 
by tuning the other parameters of the original solution to the new value of 
$\epsilon$. More precisely, at least one of the three original parameters 
[i.e. $\alpha$, $\mu$, $\nu$] should be adapted to the new value of $\epsilon$.
In this way new regular solutions can be found.
The results of the numerical 
integration and the corresponding curvature invariants are reported 
in Fig. \ref{F3} and Fig. \ref{F4} for the case $\epsilon =1$. The 
solution of Figs. \ref{F3} and \ref{F4} are obtained from the 
solution reported in Figs. \ref{F1} and \ref{F2} by adapting
the parameters according to the procedure outlined above. 
The amount of tuning can be appreciated by comparing 
the values of $\alpha$, $\mu$ and $\nu$ given in Fig. \ref{F1} with 
the ones given in Fig. \ref{F3}.

The value of $\epsilon =1 $ (corresponding to the 
integrations reported in Figs. \ref{F3} and \ref{F4} )
is only illustrative and similar solutions 
can be found for other values of $\epsilon$. Notice, however, that 
since $\mu <0 $ and $\epsilon >0$ it should always happen that 
\begin{equation}
\epsilon < \frac{5}{12 |\mu|},
\end{equation}
since only in this case the bulk solutions discussed 
in Eqs. (\ref{as1})--(\ref{as2}) are not imaginary.

Consider now the case when a given solution is divergent 
in the absence of curvature corrections. It will now be shown 
that if EGB self-interactions are included the singular 
solutions are regularized provided the value of $\epsilon$ is appropriately 
selected.

\begin{figure}
\centerline{\epsfxsize = 11 cm  \epsffile{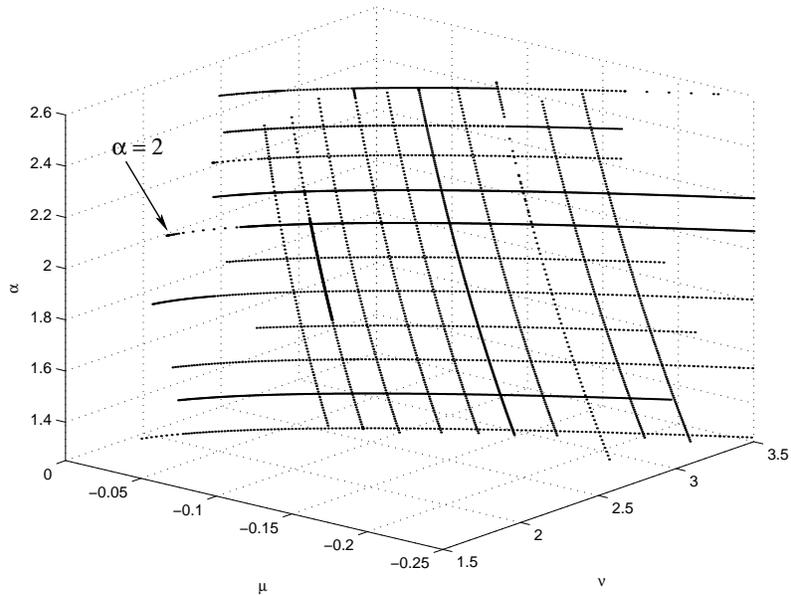}} 
\caption[a]{The parameter space of the solution in the case 
$\epsilon =0$. The curve $\alpha =2$ is also indicated.}
\label{F5}
\end{figure}

In the case $\epsilon =0$ the parameter space of the solution is represented
by a surface in the $(\alpha, \nu,\mu)$ plane. When 
the parameters lie on this surface, warped solutions 
leading to gravity localization can be found. An example 
of such a surface is given in Fig. \ref{F5}.
Let us now focus the attention on a point which lies outside 
of the fine-tuning surface. 
In this case the geometry gets singular. More specifically, two 
independent solutions can be found. These solutions 
have divergent curvature invariants for a finite value of $x$ which 
we call $x_0$. 
For $x< x_0$ the solution has two branches and can be 
parametrized as
 \begin{equation}
M(x) \sim ( x_0 -x) ^{\gamma},\,\,\,{\cal L}(x) \sim ( x_0 - x )^{\delta},
\end{equation}
with 
\begin{equation}
4 \gamma + \delta=1, \,\,\,\, 4 \gamma^2 + \delta^2 =1.
\end{equation}
These solutions 
are nothing but the well known Kasner solutions, widely discussed 
in cosmological frameworks and in the context of gravitational
collapse. 
The Kasner conditions leave open only two possibilities: either
$\delta = 1$ and $\gamma =0$ or $\gamma= 2/5$ and $\delta = -3/5$.
The branch with $\delta =1$ leads only to a coordinate singularity 
whereas the other branch leads to a curvature singularity. 

For sake of clarity consider the case when $\alpha =2$. In this case 
the parameter space of the solution $\epsilon =0$ reduces to a curve 
in the ($\mu$,$\nu$) plane  where $\alpha$ is constant 
(and fixed to $2$). This curve is indicated with an arrow in Fig. \ref{F5}.

\begin{figure}
\centerline{\epsfxsize = 11 cm  \epsffile{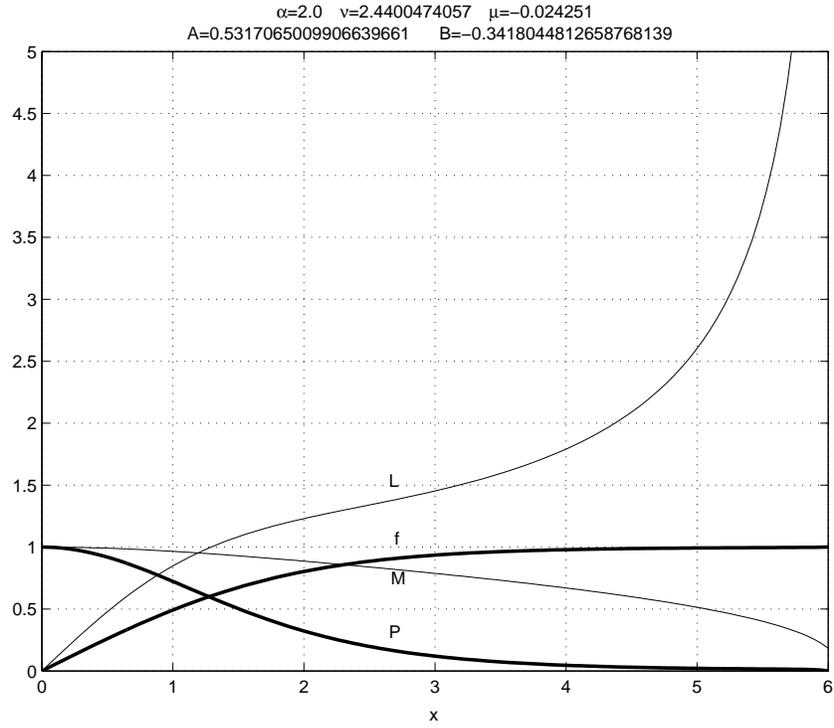}} 
\caption[a]{A singular solution outside the fine-tuning 
surface for $\epsilon=0$.}
\label{F6}
\end{figure}

\begin{figure}
\centerline{\epsfxsize = 11 cm  \epsffile{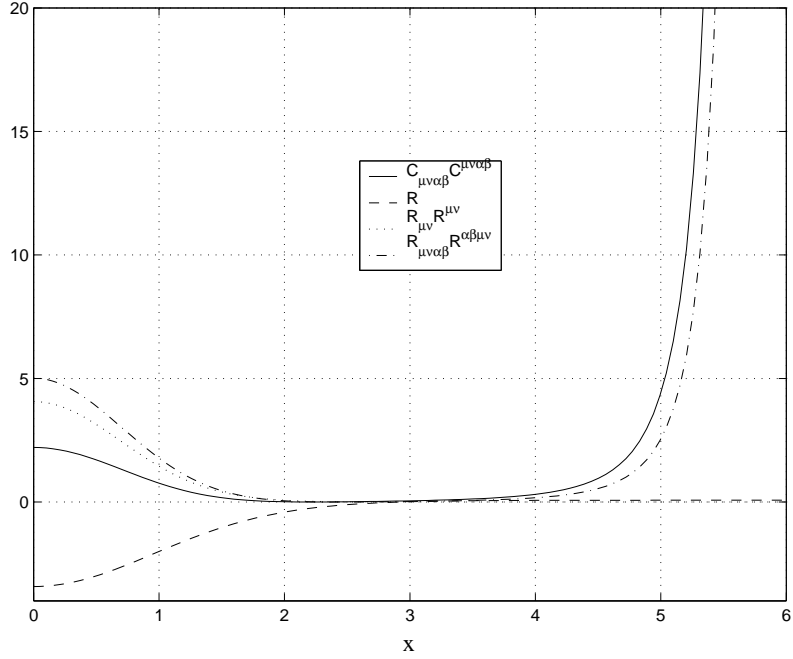}} 
\caption[a]{The curvature invariants for the solution given 
in Fig. \ref{F6}.}
\label{F7}
\end{figure}

\begin{figure}
\centerline{\epsfxsize = 11 cm  \epsffile{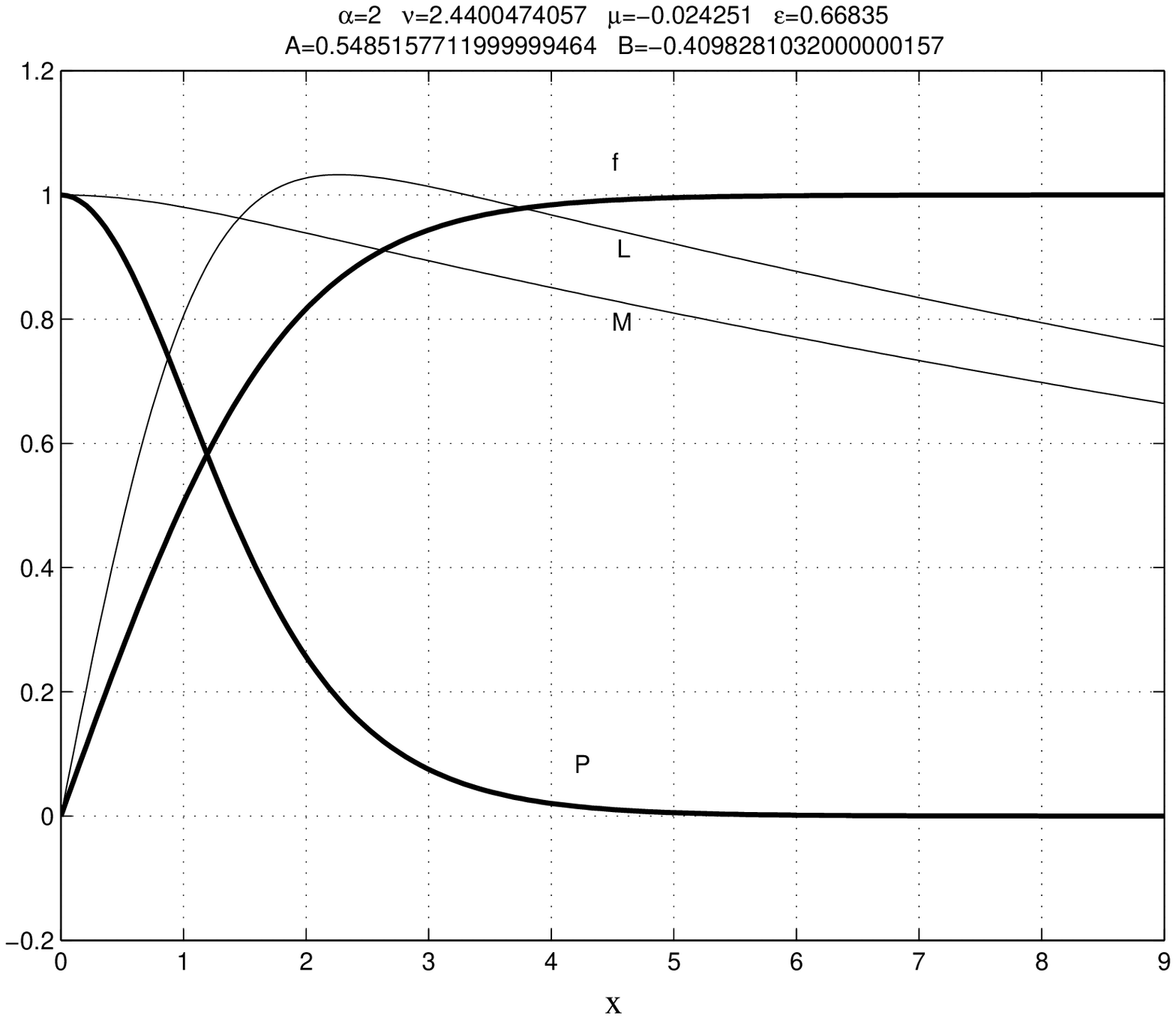}} 
\caption[a]{A regularized solution. The parameters are exactly 
the ones given in Fig. \ref{F6} except for $\epsilon$ which is 
zero in Fig. \ref{F6}.}
\label{F8}
\end{figure}

\begin{figure}
\centerline{\epsfxsize = 11 cm  \epsffile{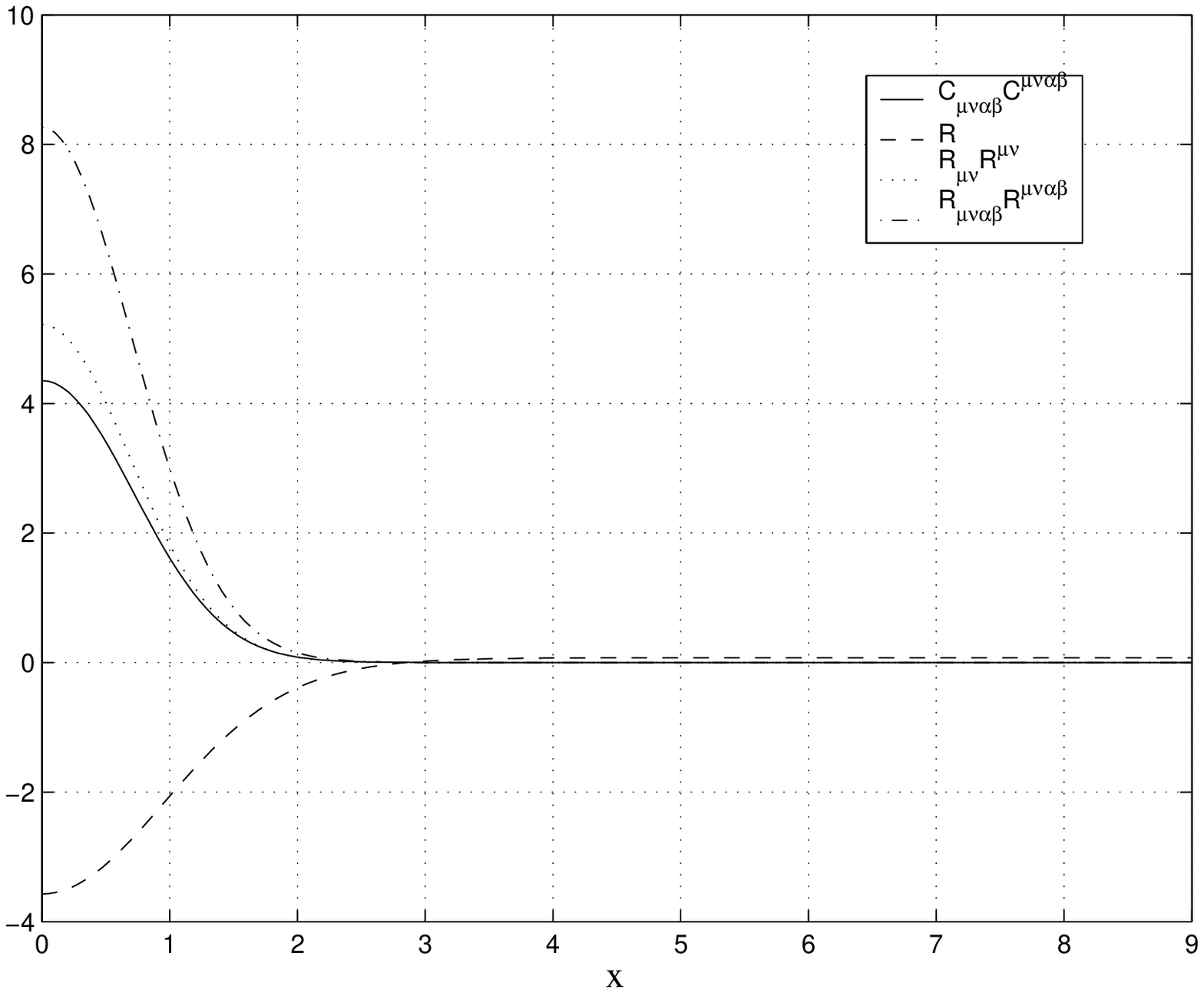}} 
\caption[a]{The curvature invariants for the solution given 
in Fig. \ref{F8}.}
\label{F9}
\end{figure}

Let us now take values of $\mu$ and $\nu$ outside of the fine-tuning 
surface in the case $\alpha=2$. Then, as argued, a singular solution appears.
In fact, in Fig. \ref{F6}, $M(x) $ goes to zero and ${\cal L}(x)$ 
blows up according to the Kasnerian relations.
The curvature invariants for the solution of Fig. \ref{F6} 
are reported in Fig. \ref{F7}. 
Figs. \ref{F6} and \ref{F7} are obtained for $\epsilon=0$. It is now possible 
to integrate the system in the case when $\epsilon \neq 0$. 
By tuning the value of $\epsilon$ 
the singular solution can be regularized. The results of this 
analysis are reported in Figs. \ref{F8} and \ref{F9}.

The curvature invariants for the solution of Fig. \ref{F7} 
are reported in Fig. \ref{F9}. 

This example is representative of a more general 
feature of the model. If a solution is singular 
for $\epsilon =0$, it can be regularized 
for by adapting the value of $\epsilon$. 
If $\epsilon \neq 0$  
the parameter space of regular solutions leading to warped compactification 
is wider than in the case $\epsilon =0$. This aspect will be 
further analyzed in Section V.

\subsection{Solutions with $\Lambda >0$.}

From the analysis of Eqs. (\ref{as1}) and (\ref{as2}), 
it is clear that regular solutions 
where $\mu >0$ and $\epsilon <0$ cannot be obtained. Hence, let us discuss
the case where $\mu> 0$ and $\epsilon>0$. We are primarily 
interested in solutions of the equations of motion describing a vortex
in a six-dimensional regular geometry with 
warp factors exponentially decreasing at infinity. A general argument 
will now show that these solutions are not possible if the cosmological 
constant is positive and $\epsilon >0$. 

If the solutions shall describe exponentially decreasing warp factors for 
large $x$, then $m(x)$ should decrease reaching, asymptotically, 
a constant negative value. This value is determined by the 
bulk solutions and it is given, in the case $\mu >0$ and $\epsilon >0$ by 
\begin{equation}
m_{\infty} = -\sqrt{\frac{1}{12 \epsilon} 
\biggl( 1 + \sqrt{ 1 + \frac{12}{5} \epsilon\mu}\biggr)}.
\label{root1}
\end{equation}
Notice now, that, from the equations of motion, a critical point 
can be defined. It is the point where the function 
$(1 - 12 \epsilon m^2)^{-1}$ 
appearing in Eq. (\ref{mbul}) blows up:
\begin{equation}
m_{s} = - \frac{1}{\sqrt{12 ~\epsilon}}.
\end{equation}
If $m(x)$ should decrease from $x=0$ towards negative values reaching, 
ultimately, $m_{\infty}$, it is 
easy to see that $m(x)$ should always pass trough $m_{s}$.
For this it is enough to notice, from Eq. (\ref{root1}), that 
in spite of the magnitude of $\mu\epsilon$ 
\begin{equation}
|m_{\infty}| > |m_{s}|.
\end{equation}
From Eq. (\ref{tuneq}) we have
\begin{equation}
(m - \ell) ( 1 - 12 \epsilon m^2) =  \frac{\nu}{\alpha} 
\frac{P P'}{{\cal L}^2}. 
\end{equation}
Hence, at the point $x_{s}$ where $m = m_s$ we will have that 
\begin{equation}
\left.\frac{P P'}{{\cal L}^2}\right|_{x_{s}} =0,
\end{equation}
and $(m(x_{s}) - \ell(x_{s}) )$ is finite. 
So, in $x_{s}$ either $P'(x_s) =0$ or $P(x_s)=0$.
Let us examine, separately, the two cases. 
Consider, first the case $P'(x_s) =0$ and assume that $m(x)$ and 
$\ell(x)$ are non singular 
around $x_s$. In this case $m(x)$ and $\ell(x)$
 can be expanded in a Taylor series 
around $x_s$. Looking at the difference between Eq. (\ref{l1})  and 
Eq. (\ref{m1}) 
\begin{equation}
( 1 - 12 \epsilon m^2) \biggl[ m ( l - m) - m'\biggr] = 
\frac{\nu}{4} \biggl( {f'}^2 - \frac{ f^2 P^2}{L^2} \biggr),
\end{equation} 
 we have that, for $x \simeq x_s$
\begin{equation}
\biggl({f_s'}^2 - \frac{f_s^2 P_s^2}{L_s^2}\biggr) =0
\label{eqfs}
\end{equation}
which also implies that $f'_s \simeq f_s P_s/L_s$. 
Looking now, separately, at Eqs. (\ref{l1}) and (\ref{m1})   we see that 
\begin{equation}
\tau_{\rho}( x_s) \simeq \tau_{\theta}(x_s) \sim
 - \frac{1}{\nu} \biggl( \mu + \frac{5}{12 \epsilon}\biggr)
= - \Omega.
\end{equation}
Since $\mu$ and $\epsilon$ are both positive, $\Omega>0$. 
But this leads us to a contradiction since we should have:
\begin{equation}
\tau_{\rho}(x_s) + \tau_{\theta}(x_s) = - 2 \Omega = 
\frac{1}{2} (f_s^2 -1)^2 - \frac{{P_s'}^2}{\alpha L_s^2}. 
\end{equation}
By hypothesis $P'_s=0$. Hence $(f_s^2 -1 )^2 = -4 \Omega$. But 
$\Omega >0$ and  $(f_s^2 -1 )^2$ is positive definite. This 
proves that $m(x)$ cannot be regular around $x_s$ since the equations 
of motion lead to a contradiction. 

Consider now the case $P_s =0$. In this case $P'_{s}$ can be either 
positive or negative. If $P'_s<0$, this means that $P(x)$ is zero in $x_s$ 
and that it decreases getting more and more negative. However, from Eq.
(\ref{boundary}), $P(x)$ should go to zero at infinity. Hence, $P'(x)$ 
should change sign for $x> x_s$. By continuity there should be a point 
$x_1$ where $P'(x_1)=0$ and $ P(x_1)<0$. From Eq. (\ref{p1}) we see that 
this would imply $P''(x_1)<0$ which is not consistent with a minimum in $x_1$.
But then, if $P_s=0$ and $P_s'<0$ the boundary values of Eq. 
(\ref{boundary}) cannot be connected continuously 
with the behaviour in the core.
A similar demonstration can be obtained in the case $P_s=0$ and  $P'_s >0$. 

\renewcommand{\theequation}{6.\arabic{equation}}
\setcounter{equation}{0}
\section{The parameter space  of the solutions} 

In the previous Section we discussed examples where 
singular solutions are regularized in the presence of the EGB 
term. In this Section the parameter space of the solution 
will be discussed. It will be shown how the presence 
of quadratic corrections allows to enlarge the parameter space. 

\begin{figure}
\centerline{\epsfxsize = 11 cm  \epsffile{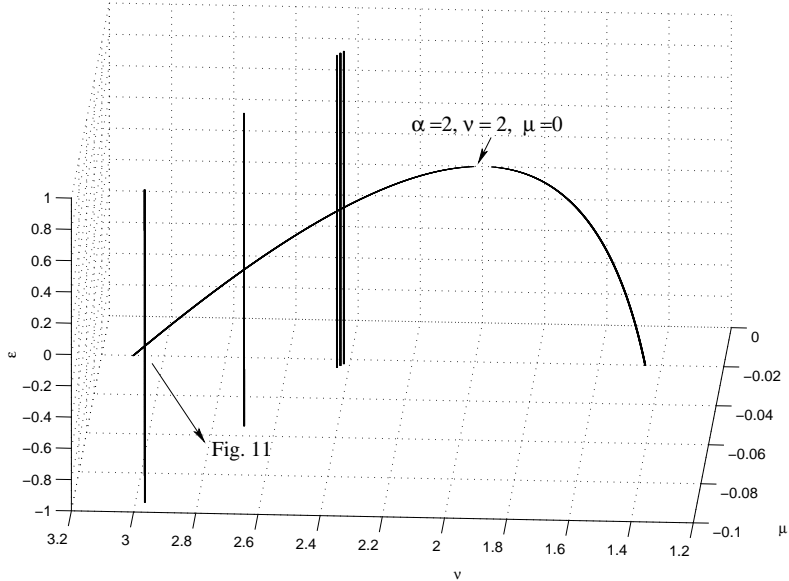}} 
\caption[a]{The parameter space of the solution is illustrated 
for the case $\epsilon \neq 0$.}
\label{F10}
\end{figure}

\begin{figure}
\centerline{\epsfxsize = 11 cm  \epsffile{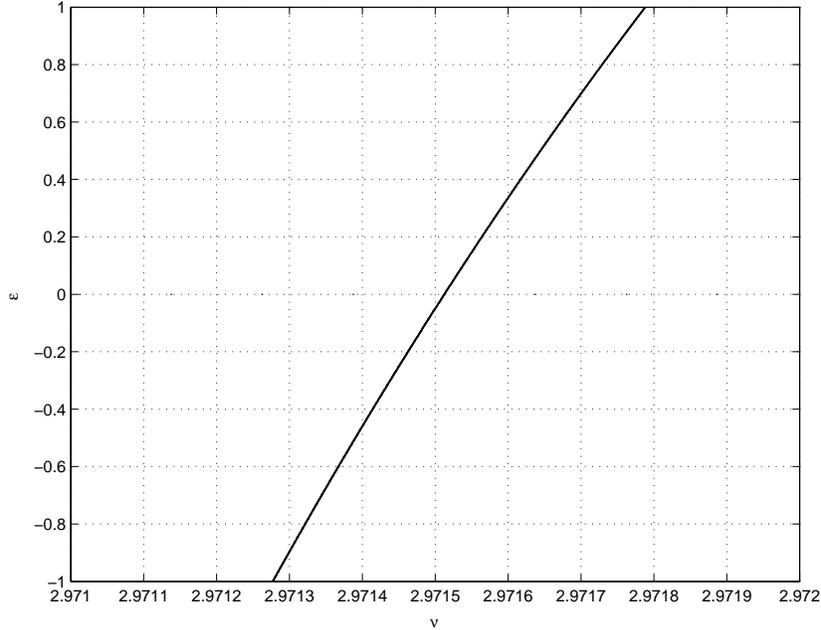}} 
\caption[a]{The projection, in the $(\epsilon,\nu)$ plane, 
 of the curve with largest $\mu$ appearing in Fig. 
\ref{F10}.}
\label{F11}
\end{figure}

In order to do this let us fix one of the three parameters $(\alpha, 
\mu, \nu)$ and let us see what happens to the parameter space 
when $\epsilon \neq 0$. Suppose, for instance, that $\alpha$ is fixed.
 Consider, for consistency with the examples of Secion IV, 
the case when $\alpha =2$.
This case corresponds to the Bogomol'nyi limit \cite{bog}.  
In the case $\epsilon =0$ the fine-tuning curve is illustrated 
in Fig. \ref{F5}. In Fig. \ref{F10} the parameter space is 
studied in the case $\epsilon \neq 0$ for $\alpha = 2$. 
The ``parabola'' appearing in Fig. \ref{F10} represents the 
parameter space of the solution for $\alpha =2$ and $\epsilon =0$. 
On the {\em almost} vertical lines, $\epsilon$ changes. 
The lines appearing in Fig. \ref{F10} are, indeed,  not truly vertical. This 
aspect is illustrated in Fig. \ref{F11} where the 
projection of one of the lines of Fig. \ref{F10} is reported 
for constant $\mu$. 
The points slected outside the ``parabola'' would lead, in the 
absence of quadratic corrections,
to singular solutions, as discussed in Section IV. However, if 
quadratic corrections are present, the parameters of the solution 
can be adapted to the non-vanishing value of $\epsilon$. 
In this way,  new regular solutions can  be found 
 if $\epsilon$ stays on the curves intercepting 
the ``parabola'' lying, in its turn, on the $(\mu, \nu)$
plane.  

From a practical point of view the various curves intercepting the 
``parabola'' have been obtained in the following way.
Starting from the point on the ``parabola'', $\mu$ was held constant, 
$\epsilon$
was varied by small steps and $\nu$ was tuned so that the
relation $B=-nu/2alpha$ stays satisfied.

The possibility of finding these regular solutions 
means that the parameter space is wider in the case when $\epsilon \neq 0$ 
than in the case when $\epsilon =0$. The same kind of analysis can be 
numerically 
performed starting from any point lying on the ``parabola'' and the 
five curves reported are only illustrative.
As previously stressed, the curves intercepting the ``parabola'' cannot be 
extended to arbitrary large values of $\epsilon >0$. In fact, since $\mu <0$ 
and $\epsilon >0$, from Eqs. (\ref{as1})--(\ref{as2}) we must have that 
$\epsilon < 5/ (12 |\mu|)$ in order to avoid imaginary warp factors.

In Fig. \ref{F11} also positive values of $\mu$ have been included. 
As discussed in the previous Section, expoenentially decreasing 
solutions for the warp factors are not possible if $\mu$. 
Moreover, also in the case 
$\epsilon=0$ exponentially decreasing warp factors are not allowed. The part
of the curve corresponding to $\mu>$ leads, for $\epsilon=0$ to
exponentially {\em increasing} warp factors, as discussed in \cite{us}. 
The border point between the regions $\mu<0$ and  $\mu>0$ 
of the parameter space is represented by the  Bogomol'nyi point 
where $\mu=0$, $\nu =2$ and $\alpha =2$ \cite{us}.

In the case when quadratic curvature corrections are absent the 
parameter space of the model is three-dimensional and it can be described 
with a three dimensional plot as in Fig. (\ref{F5}). If quadratic curvature 
corrections are present the parameter space is four-dimensional 
and it has to be studied, necessarily, one of the four parameters. In the 
present discussion $\alpha$ has been fixed. However, the same 
study can be repeated by fixing either $\mu$ or $\nu$ and by discussing, 
respectively, the picture of the parameter space in the ($\alpha$, $\nu$, 
$\epsilon$) or ($\alpha$, $\mu$, $\epsilon$) planes.

\renewcommand{\theequation}{7.\arabic{equation}}
\setcounter{equation}{0}
\section{Concluding remarks} 

In this paper the solutions of the gravitating Abelian-Higgs model 
in six-dimensions have been studied. Provided the quadratic 
corrections are parametrized in the EGB form, new classes of regular 
solutions leading to warped geometries have been found. 
The rationale for these results stems from the fact that  the
relations among the string tensions still exist even if 
curvature self-interactions are present. In its turn, 
this property, relies, ultimately, 
on the features of the EGB combination which does not produce, in the
equations of motion, derivatives higher than second.

If the bulk cosmological constant is negative the results of adding
 the EGB term in the bulk action can be viewed in two 
different perspectives. Starting from a regular solution (without quadratic 
corrections), new solutions (still regular) can be found in the
vicinity of the initial solution by changing 
$\epsilon $ from $0$ to $\epsilon > 0$.
If, in another perspective, we start from a singular solutions 
(without quadratic corrections) new (regular) solutions can be found by 
tuning appropriately the value of $\epsilon$.
  
The results obtained in this investigation show that vortex-like 
solutions,  obtained using the Einstein-Hilbert action,  are stable 
towards the introduction of quadratic corrections of the EGB form. 
Furthermore, 
our results suggest that in  the parameter space of the regular 
solutions a new direction opens up when $\epsilon \neq 0$.

The singularity-free solutions 
discussed in the present paper represent an ideal framework in order 
to check for the localization of the various modes of the geometry 
as discussed in \cite{mg4,mg5}. The presence 
of an (Abelian) gauge field background makes these solutions also 
rather interesting from the point of view of fermion localization \cite{tro}.
These problems are left for future studies. 

\section*{Acknowledgements}

The authors are deeply indebted to M. Shaposhnikov for inspiring 
discussions and important comments. The support of the Tomalla foundation 
is also acknowledged.

\end{document}